\documentclass[aps, prx, reprint, longbibliography, nofootinbib,superscriptaddress, floatfix]{revtex4-2}
\RequirePackage{snapshot}
\usepackage{slashed}
\usepackage{verbatim}
\usepackage[T1]{fontenc}
\usepackage{mathbbol}
\usepackage[dvipsnames]{xcolor}
\usepackage{orcidlink}

\usepackage[normalem]{ulem}
\usepackage[english]{babel}
\usepackage{lipsum}
\usepackage{physics}
\usepackage{dcolumn}
\usepackage{tensor}
\usepackage{comment}
\usepackage{placeins}
\usepackage{graphicx,color,overpic,mathtools}
\usepackage{amsthm,amsmath,amssymb,mathrsfs}
\usepackage{braket,bm,bbm,setspace}
\usepackage{booktabs}
\usepackage{cancel}
\usepackage{float}
\usepackage{xargs}
\definecolor{myred}{RGB}{179, 27, 27}
\usepackage{hyperref}
\hypersetup{
    colorlinks=true,
    linkcolor=myred, 
    citecolor=myred, 
    urlcolor=myred  
 }

\newcommand{\ie}{i.\,e.\@}

\begin{document}

\title{The weight of light:
colliding pulses of radiation in General Relativity
}

\author{Jorge Exp\'osito Patiño\orcidlink{0009-0007-5136-3126}} 
\affiliation{CENTRA, Departamento de F\'{\i}sica, Instituto Superior T\'ecnico -- IST, Universidade de Lisboa -- UL, Avenida Rovisco Pais 1, 1049-001 Lisboa, Portugal}

\author{Vitor Cardoso\orcidlink{0000-0003-0553-0433}}
\affiliation{Center of Gravity, Niels Bohr Institute, Blegdamsvej 17, 2100 Copenhagen, Denmark}
\affiliation{CENTRA, Departamento de F\'{\i}sica, Instituto Superior T\'ecnico -- IST, Universidade de Lisboa -- UL, Avenida Rovisco Pais 1, 1049-001 Lisboa, Portugal}

\author{Hannes R. Rüter\orcidlink{0000-0002-3442-5360}}
\affiliation{CENTRA, Departamento de F\'{\i}sica, Instituto Superior T\'ecnico -- IST, Universidade de Lisboa -- UL, Avenida Rovisco Pais 1, 1049-001 Lisboa, Portugal}

\begin{abstract}
We study the gravitational interaction of very strong pulses of 
electromagnetic radiation, up to and beyond gravitational collapse. 
We demonstrate the existence of very compact states close to the
threshold of collapse, both for a single pulse, and a head-on collision of 
such pulses, and
we measure their compactness to be above the maximum value proposed by the hoop
conjecture.
For a single pulse above the threshold of collapse,
a fast black hole is produced.
Depending on the amount of tuning, head-on collisions can form two  horizonless states,
two small black holes or one big black hole.
Our results show that the gravitational interaction of massless fields can
lead to nontrivial final or intermediate states,
while giving rise to extremely high luminosity and compactness.
\end{abstract}
\maketitle

\noindent\textbf{\emph{Introduction.}}
High-energy collisions are an irreplaceable tool in particle physics.
The study of the end products from the splashing of nuclei or elementary
particles against each other shaped our view of the Standard Model.
Higher and higher energies are key to probing smaller and smaller scales,
where new physics might emerge.
General Relativity introduces an interesting twist to this story.
At transplanckian center of mass energies, the collision should form a black
hole (BH) that is large with respect to the rest masses of the
particles~\cite{Banks:1999gd,Giddings:2001bu,Sperhake:2012me,Cardoso:2014uka},
effectively hiding from view all structure of the colliding objects.
This is encapsulated in the paradigm%
-- anchored also on the hoop conjecture~\cite{Thorne:1982cv, Flanagan} -- that ``matter does not matter'' at high energies and that BHs truly represent
the end of particle physics~\cite{Giddings:2001bu,Sperhake:2012me,
Cardoso:2014uka}.

Ultrarelativistic collisions have been studied
analytically~\cite{Szekeres:1970vg, Khan:1971vh, Penrose:1974, Luk:2013zr,
Eardley:2002re}
and numerically~\cite{Sperhake:2008ga,Sperhake:2009jz,Shibata:2008rq,
East:2012mb,Healy:2015mla,Healy:2024lhl,Bozzola:2022uqu,Zhu:2026mhn}.
Analytical studies can only describe the resulting spacetime for a short time
after the collision,
and numerical works use massive colliding objects,
which are notably difficult to boost to very high speeds.
This is because of the disparity of length scales between the
Lorentz-contracted colliding objects and the very massive remnant.
Moreover, to prevent spurious radiation in the initial data from significantly
affecting the result, very large separations are needed.

Here, we investigate the ultrarelativistic regime by studying
collisions of light (\ie{} pulses of focused electromagnetic radiation)
with amplitudes that are strong enough for light to significantly self-gravitate.
Considering a single such focused pulse, which moves at the
speed of light: is there a critical amplitude beyond which it can
collapse to a BH?
If so, is the Hoop conjecture still a faithful guide?
What is the speed of such BH?
Through nonlinearities,
can we transfer significant energy to very small scales?
When two such pulses collide,
is it possible for the EM energy to self-gravitate,
forming bound configurations?

The collapse of EM fields has been studied numerically in the context of
critical collapse~\cite{Baumgarte:2019fai,Mendoza:2021nwq,Reid:2023jmr}
for initial conditions that do not correspond to incoming radiation from
infinity (since the initial data are set up at the focusing point directly).
The traditional picture of critical collapse
focuses on tracking maximum variables and apparent horizon mass, which
we extend to fully characterize the physics of matter
so strongly self-gravitating.
In particular,
those methods do not probe the presence of very compact horizonless states
close to the threshold of collapse.
Asymptotic quantities such as gravitational waveforms are crucial for providing
a full picture.

\begin{figure*}
\centering
\includegraphics[width=0.9\linewidth]{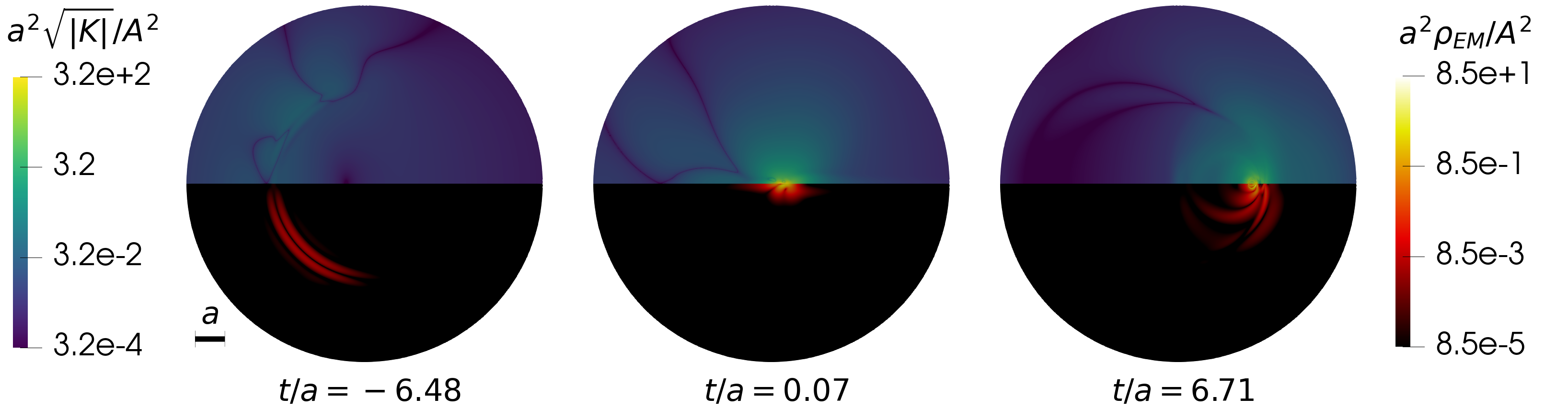}
\caption{
    \label{fig:pulses}
  Evolution of a single pulse for a close-to-critical amplitude $A = 0.5271$.
  The disks are sections through the $xz$ plane and have a radius of $12a$.
  The horizontal ($z$) axis divides the disks into halves.
  The top halves show the square root of the Kretschmann scalar,
  and the bottom halves show the EM energy density. Both are rescaled by $A^2$,
  which is the scaling of the leading term at low amplitudes.
  In a linear evolution, the pulse would focus at $t=0$.
  }
\end{figure*}
\noindent\textbf{\emph{Framework.}}
We use geometric units $(G = c = 1)$ and Gaussian units for the Maxwell
equations $(4\pi\varepsilon_0 = 1)$ throughout. Greek indices range from
0 to 3 and Latin indices range from 1 to 3.

We study electrovacuum solutions, governed by the Einstein-Maxwell system,
\begin{gather}
  R_{\mu\nu} - \frac 12 R g_{\mu\nu} = 8\pi T_{\mu\nu} \, , \\
  \label{eq:maxwell_F}
  \nabla_\mu F^{\mu\nu} = 0\,,\quad \nabla_\mu (F^*)^{\mu\nu} = 0 \,, \\
  T_{\mu\nu} = \frac 1 {4\pi} \bigg(F_{\mu\lambda} F_{\nu}{}^{\lambda} 
    - \frac 1 4 g_{\mu\nu} F_{\rho\sigma} F^{\rho\sigma}\bigg) \,,
\end{gather}
with $F_{\mu\nu}$ the Faraday tensor and
$(F^*)^{\mu\nu} = -\frac12 \epsilon^{\mu\nu\rho\sigma} F_{\rho\sigma}$ its Hodge
dual.
We use the convention that $\epsilon^{0123} = -1/\sqrt{-g}$ and
$\epsilon_{0123} = + \sqrt{-g}$~\cite{Alcubierre:2009ij}.
We perform a 3+1 split of the geometry.
Introducing the the timelike normal $n^\mu$ to the spacetime
sheets we can express the EM components as
\begin{gather}
  F^{\mu\nu} = n_\sigma \epsilon^{\sigma\mu\nu\rho} B_\rho + n^\mu E^\nu
  - n^\nu E^\mu\,.
\end{gather}
$E^\mu$ and
$B^\mu$ are the electric and magnetic fields measured by Eulerian
observers, both spatial. We use adapted coordinates such that $E^i$ and $B^i$ can be written with spatial indices only.
Details on the 3+1 formalism can be found in
textbooks on numerical relativity~\cite{Alcubierre:2009ij,AlcubierreBook,
Baumgarte:2010ndz,Gourgoulhon:2007ue}.

In the initial data,
we specify conformal EM fields that are solutions of the linear EM constraints.
By construction,
the initial data automatically satisfy the EM constraints of the fully
nonlinear system,
see~\cite{Alcubierre:2009ij,Patino:2025epy}.
To satisfy the Einstein constraints, we numerically solve the
conformal-thin-sandwich equations~\cite{Baumgarte} for the conformal
factor and shift.

We construct the conformal EM fields with the flat space method of the 
Hertz potentials~\cite{VARGA1998108}~\cite[pp.~281--282]{Jackson:1998nia}.
Given a Hertz potential $\Pi_k$, we construct the EM potentials of the
flat conformal geometry by
\begin{gather}
  (\tilde \varphi, \tilde A^i) = (0, \epsilon^{ijk} \partial_j \Pi_k) \,.
\end{gather}
Our initial data corresponds to a nonlinear extension of the localized causal
pulses introduced in~\cite{Lekner2018ElectromagneticPL}, 
whose Hertz potential is of the form
\begin{gather}
  \Pi_k = \tilde c A \delta^z_k \text{Re}\left[\frac{a^5(a + it_0)}{3}
  \frac{3\bar\rho^4 - 6 z^2 \bar\rho^2 - z^4 + 8 i z \bar\rho^3}
  {\bar\rho^3(\bar\rho^2 + z^2)^3}
  \right]\,,
\end{gather}
with $\bar\rho^2 = x^2 + y^2 + (a+it_0)^2$,
$x$, $y$ and $z$ Cartesian coordinates,
and $t_0$ is the time at which we solve the constraints, with $t_0 = 0$
corresponding to the focus time.
The parameter $a$, with units of length, sets the overall scale of the 
solution and $A$ is a dimensionless amplitude parameter.
For convenience, the constant $\tilde c$ is chosen such that in the linear
regime, the ADM mass is given by $M_{\rm ADM}/a = A^2$.
Our results are invariant under appropriate rescaling with $a$.
For reference, in SI units, an electric field of magnitude $1/a$
corresponds to
$E =1.0429\left(\frac{1\,\text{m}}{a}\right)\times10^{27} \,\text{V/m}$
and an energy density of magnitude $1/a^2$ is 
$\rho_{\rm EM} = 1.21\left(\frac{1\,\text{m}}{a}\right)^2 \times 10^{44}\,\text{J/m}^3$.
For a beam waist size of one 1~mm, this is about 14 orders of magnitude above
quark-gluon plasma energy densities at the Large Hadron Collider,
which are around $10^{36}$~J/m$^3$~\cite{Busza2018}.

The first panel of Fig.~\ref{fig:pulses} shows an early snapshot of the
solution,
effectively in the linear regime.
The pulse is symmetric under rotation around the z-axis (horizontal in Fig.~\ref{fig:pulses}),
is localized within a small radial shell
and propagates in the direction of the positive z-axis.
The focusing region has a width of a few $a$.
In the asymptotic past, these pulses are spread out, resulting in very weak
self-gravitating effects.
For all initial data, we use a time parameter of $t_0=-8a$, which is enough 
to start the simulation effectively in the linear regime. Only as the pulse moves towards the origin do the nonlinear effects become
significant.

The hoop conjecture states that a BH forms if the compactness of matter
reaches one~\cite{Flanagan, Thorne:1982cv}
\begin{gather}
  C = \frac{2M}{R} = 1\,.
  \label{eq:hoop}
\end{gather}
There are several similar trapped surface formation statements,
both conjectured and proven~\cite{Schoen:1983tiu,Hirsch:2023grt,Bizon:1989xm}.
For our purposes, the mass $M$ in a region is measured by Eulerian observers
by integrating $\rho = T^{\mu\nu} n_\mu n_\nu$.
Other versions also consider the momentum measured by Eulerian observers,
in such a way that plane waves and null fluids have zero
mass~\cite{Schoen:1983tiu,Hirsch:2023grt}.
Extrapolating from the linear regime,
the hoop conjecture predicts that the critical amplitude would be $A_*\simeq
1.34$.
Extrapolation from the linear regime must necessarily be an overestimation of
the critical amplitude because the nonlinear effects involve self-gravitation of the
electromagnetic energy density,
resulting in increased compactness for any given mass~\cite{East:2012mb}.
Furthermore, extrapolation from the linear regime would indicate the presence of
a mass gap,
at the threshold of collapse BHs would have masses of order $a$.
The literature on critical collapse~\cite{Gundlach:2025yje} finds no such mass
gap for massless fields.

We evolve the Einstein-Maxwell system with
\textsc{BAMPS}~\cite{Hilditch:2015aba, Renkhoff:2023nfw}, an
MPI-parallel multigrid pseudospectral code for the generalized harmonic
gauge formulation of GR.
A description of the Einstein-Maxwell implementation can be found in
Ref.~\cite{Patino:2025epy}.
The parameter files for our simulations can be found in the data
release~\cite{ZenodoData},
and a resolution comparison can be found in the Supplemental Material.

\noindent\textbf{\emph{Single Pulse.}}
\begin{figure*}[ht!]
\centering
  \includegraphics[width=0.32\linewidth]{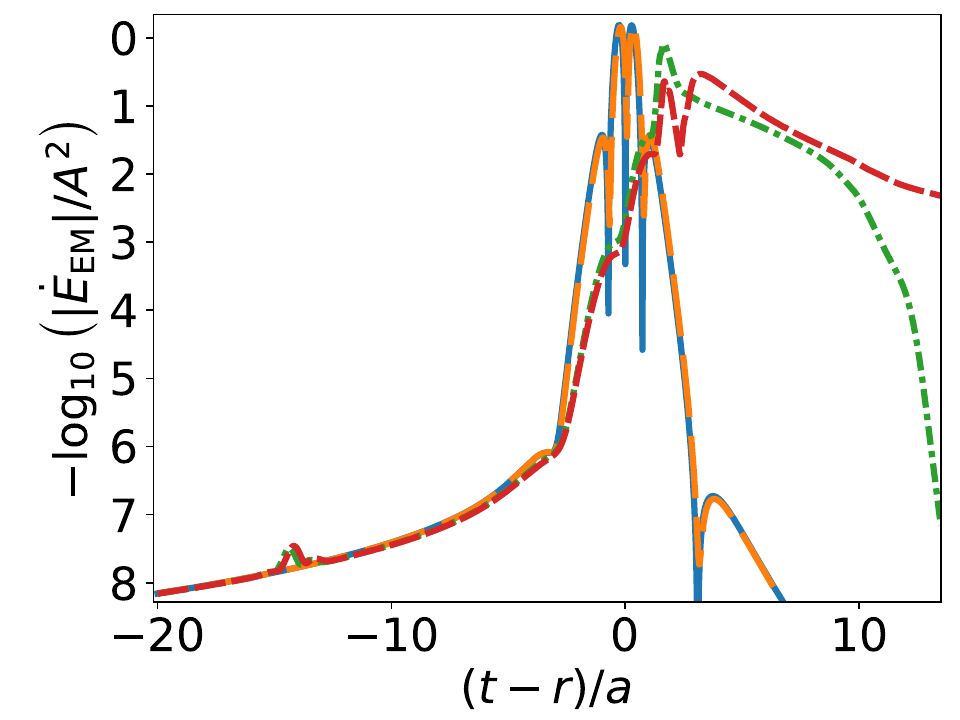}
  \includegraphics[width=0.32\linewidth]{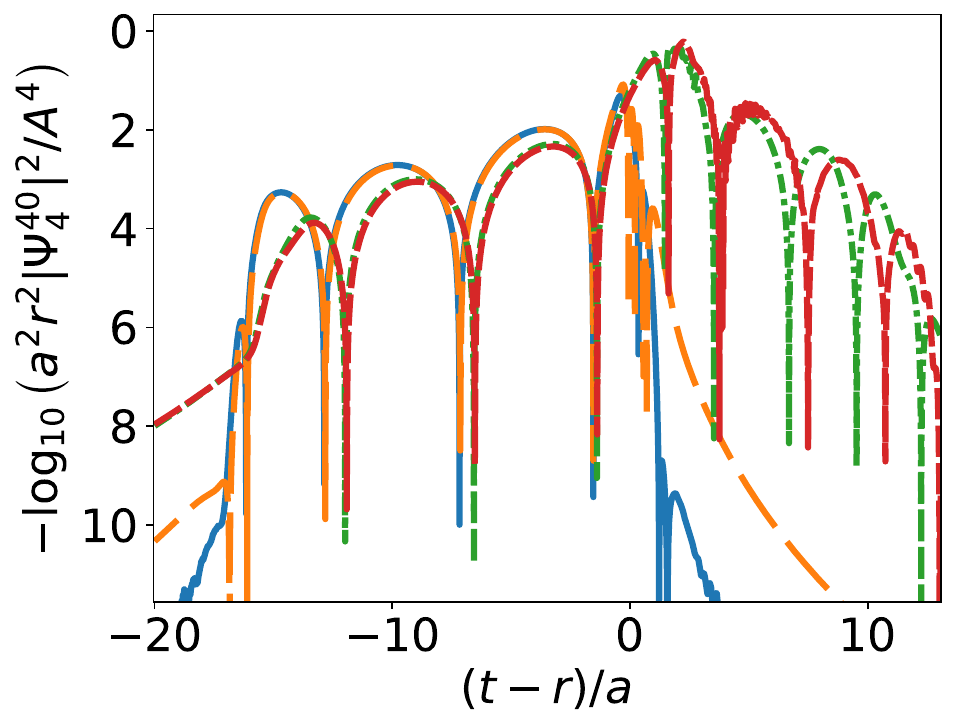}
  \includegraphics[width=0.32\linewidth]{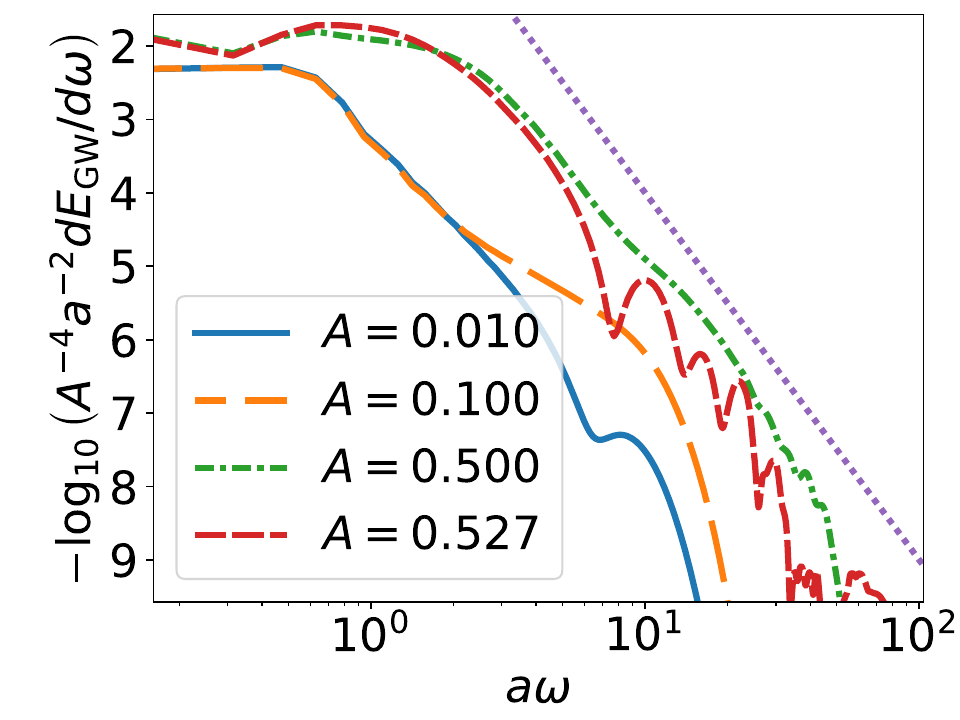}
  \caption{
  \label{fig:comparison_gws_pulse}
  Radiation properties during the evolution of a single EM pulse, for different amplitudes.
  All of the plots are rescaled by the expected leading order,
  so that in the small amplitude regime the waveforms coincide.
  {\bf Left:} The total flux of EM radiation, crossing a sphere of radius
  $r=90 a$.
  {\bf Middle:} Mode $(l, m) = (4, 0)$ of $\Psi_4$.
  {\bf Right:} Power spectrum of gravitational wave emission,
  from all of the modes up to $l=12$.
  Since the compact state is moving fast in the positive $z$ direction,
  beaming becomes important~\cite{Boyle:2015nqa},
  and there might be a relevant contribution from higher modes that we are
  neglecting.
  The spectrum also shows a best fit power law of the form $\omega^{-5}$ as a
  reference.
  }
\end{figure*}
We find that for amplitudes above a critical threshold $A_* \simeq 0.5271$,
the evolution of the initial data leads to BH formation.
This threshold is within a factor of three smaller than the expected value from linear extrapolation
using the hoop conjecture. After formation, the BH moves fast and expands. For the $A = 0.528$ simulation,
the coordinate speed of the coordinate center of the trapped region is 0.64.

Closer to the threshold of collapse,
the coordinate time until horizon formation, and its spatial location, increase,
whereas the horizon mass decreases.
This strongly suggests that a zero mass BH forms at the threshold.
Thus, unlike generic arguments based on the hoop conjecture and plane waves would
suggest,
there is no indication of a mass gap,
in line with the traditional picture of critical collapse~\cite{Gundlach:2025yje}.
In section~\ref{section:apparenthorizons} of the Supplemental Material,
we provide more details on the first detected apparent horizons.

Figure~\ref{fig:pulses} shows snapshots of the subcritical simulation with
amplitude $A = 0.5271$,
which is close to the threshold of collapse.
Nonlinear effects are very pronounced, which is reflected in the fact
that the simulation is not symmetric under the transformation $(t\rightarrow-t,z\rightarrow-z)$ (as it would in a linear evolution),
but instead the self-gravity strongly holds the pulse together and delays dispersion.

The literature on critical collapse places a special emphasis on tracking horizon mass and curvature scalars, which we find is insufficient to fully characterize the physics of solutions close to the threshold.
Based on local compactness and radiative properties, we find that
close-to-critical solutions are better understood as very compact
states with a stability time that increases as we get closer to the threshold of
collapse,
as measured by asymptotic observers.

We infer the presence of compact states from the radiation properties in the
subcritical solutions,
summarized in Fig.~\ref{fig:comparison_gws_pulse}.
For $t-r>0$,
the Weyl scalar $\Psi_4$ shows a ``ringdown'' with a frequency that decreases as
one approaches the critical amplitude.
The quality factors of these modes are lower than those of a nonspinning
BH~\cite{Berti:2009kk,Berti:2025hly}.
We attribute the difference to the presence of matter around the expected
light-ring which,
in terms of the eikonal approximation,
would result in a higher Lyapunov exponent and lower quality factors.
The decreasing ringdown frequency can be interpreted as a slight increase in
radius of the compact state.
We provide additional evidence for the interpretation of the waveforms as
ringdown of a compact state in the Supplemental Material,
where we show that they conform to a eikonal behavior (Section
~\ref{section:eikonal} of the supplemental material).
From the eikonal approximation,
we get a rough estimate on the radius of the light ring,
which is $r_\mathrm{light} = 3.2 a$ in the $A=0.5$ simulation.

The EM waveforms also support the presence of a compact state.
We calculate the luminosity $\dot{E}_{\rm EM}$ in the EM channel as:
\begin{gather}
\dot{E}_{\rm EM}\equiv \frac{dE_{\rm EM}}{dt} = \lim_{r\to\infty} \frac{r^2}{4\pi} \oint |F_{\mu\nu} \bar m^\mu k^\nu|^2
 d\Omega\,,
\end{gather}
with $k^\mu$ and $m^\mu$ part of a Newman-Penrose tetrad
$\{l^\mu, k^\mu, m^\mu, \bar m^\mu\}$, $k^\mu$ being the ingoing null vector.
The EM luminosity drops quickly after a few $a$ for low amplitude pulses.
However, close to the threshold,
the luminosity decay is prolonged over increasingly longer time spans,
again suggesting it is trapped within a compact geometry, and slowly leaking out.
Notably, in the $A=0.5$ simulation, the luminosity reaches a maximum of $0.21$,
reaching close to the conjectured upper limit~\cite{Cardoso:2018nkg}.

We complement the gauge independent asymptotic study by estimating the
compactness measured by Eulerian observers close to the center of the
simulation domain, as described in the Supplemental
Material~\ref{section:compactness}.
Our measurement does not correspond exactly to spheres in the slice geometry,
which are the regions considered in the hoop conjecture.
That said, we consistently find compactness over one for ellipsoids of different
radii and different centers along the z-axis for simulations close enough to the
threshold.
In particular, for a simulation with $A = 0.5271$, at time $t = 2.23a$,
we find a region of extremely high compactness of 2.32 centered at $z=2.17a$.
This shows that Eulerian observers see the presence of a very compact state
close to the threshold of collapse.

\begin{figure*}[ht!]
  \centering
  \includegraphics[width=0.9\linewidth]{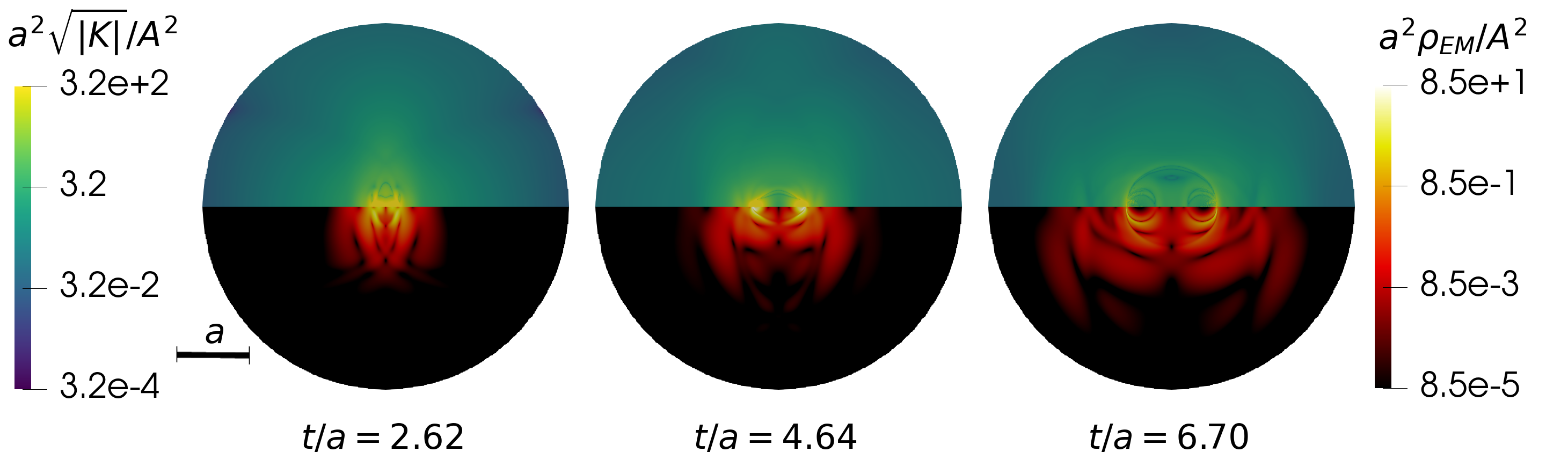}
  \caption{\label{fig:headon}
  Same setup as Fig.~\ref{fig:pulses} but for a head-on collision of subcritical pulses
  with an amplitude of $A = 0.38$.
  These figures are centered at $z = 0$, and have radius $5a$ in order to show
  the transition from a single compact state to two, to dispersion.
  The snapshot times differ from Fig.~\ref{fig:pulses} to display
  the more relevant features.
  }
\end{figure*}
\noindent\textbf{\emph{Head-on collisions of pulses.}}
We turn our attention to head on collisions of such pulses.
Now, the initial data is constructed with the same method outlined above,
with conformal EM data corresponding to the superposition of two pulses moving
towards each other.
The focus positions of the pulses are set up such that, in the linear regime,
they would have a separation of $2a$.

Apparent horizons form when the amplitude is $A\gtrsim 0.383$.
Collisions with amplitudes substantially higher than the threshold result in a
single BH at the center that covers both pulses.
We interpret the collapse in these cases to be triggered by the interaction of
the two pulses.
As we approach the threshold,
two distinct BHs form after the collision and initially move away from each
other.  
For the amplitude $A = 0.384$,
we use a Newtonian estimate to show that the BHs likely are bound and would
merge at a later point.
The BHs have an initial coordinate velocity of $v = 0.1$,
from which we derive an approximate kinetic energy of $0.0006 a$, much smaller
than the Newtonian gravitational potential, $m^2/(z_1 - z_2)=0.013 a$.
This type of behavior also occurs in ultrarelativistic fluid
collisions~\cite{East:2012mb}.
As we tune closer to the threshold of collapse,
we find that the distance between apparent horizons increases,
although with the current results we cannot conclude if it reaches a finite
value or it increases without bound.
Section~\ref{section:apparenthorizons} of the Supplemental Material
summarizes the properties of the first
detected horizons for a few supercritical simulations.

\begin{figure}[ht!]
  \centering
  \includegraphics[width=0.95\linewidth]{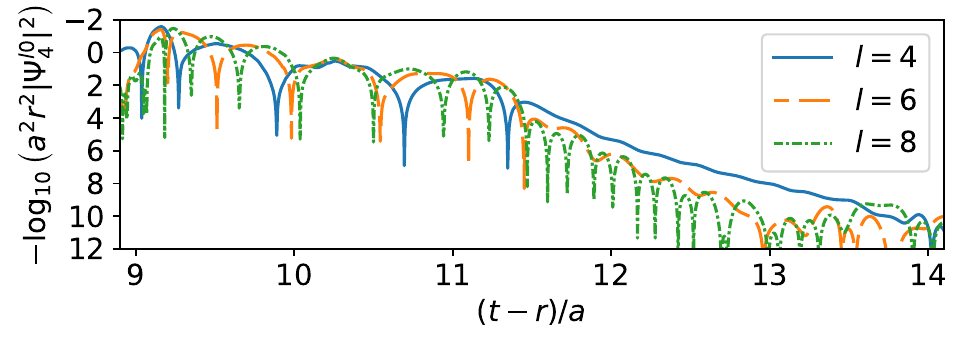}
  \caption{
  \label{fig:double_ringdown}
  Spin-weighted spherical harmonic mode contributions of $\Psi_4$ for a
  subcritical head-on simulation with $A=0.381$,
  extracted at $r=90a$.
  }
\end{figure}

On the subcritical side,
solutions close to the threshold first form a single compact state that splits
into two more compact ones,
as shown in snapshots of the energy density and Kretschmann scalar in Fig.~\ref{fig:headon}.
For the simulation with $A=0.38$,
the first state has compactness 0.77 and the independent states have compactness
1.09.

This local picture is supported by the $\Psi_4$ waveforms,
shown in Fig.~\ref{fig:double_ringdown},
which for higher modes display a double ringdown,
the second one with more than double the frequency of the first.
This is in contrast to the single pulse case,
where there is a single ringdown phase.
We interpret the first ringdown phase as stemming from the
emission from a single coalesced state,
whereas the second phase originates from the two individual compact states exiting the interaction region.
The second ringdown part of the $l=4$ mode in our signal has very small oscillations on
top of a purely decaying mode.
This behavior is compatible with the results in~\cite{Capuano:2026tjy} which show that
accreting or radiating spacetimes may possess purely decaying modes.

\begin{figure*}[ht!]
  \centering
  \includegraphics[width=0.32\linewidth]{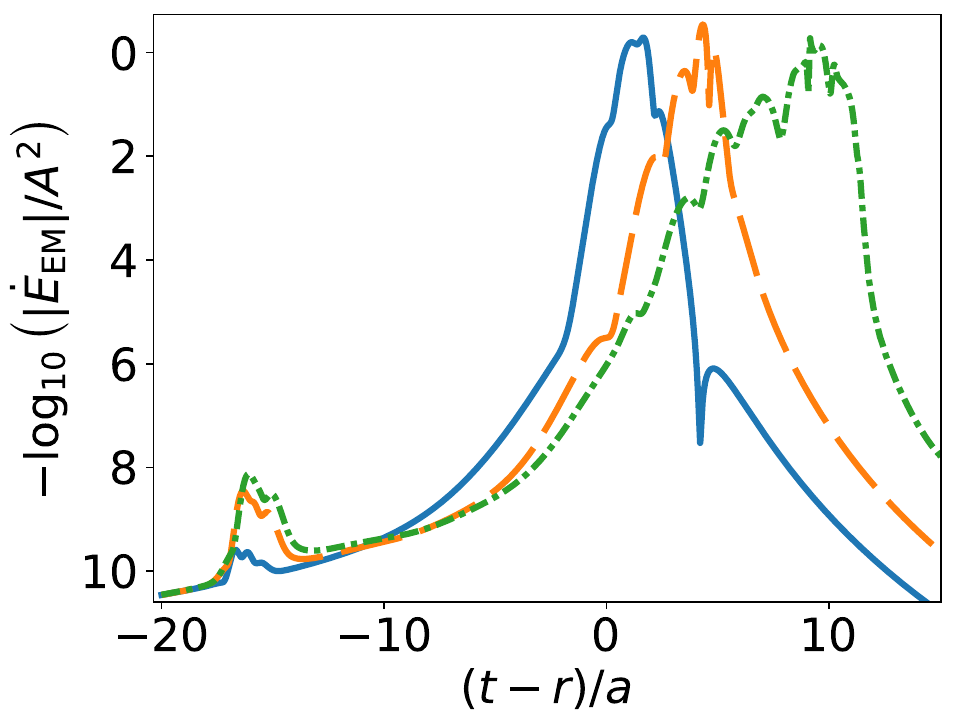}
  \includegraphics[width=0.32\linewidth]{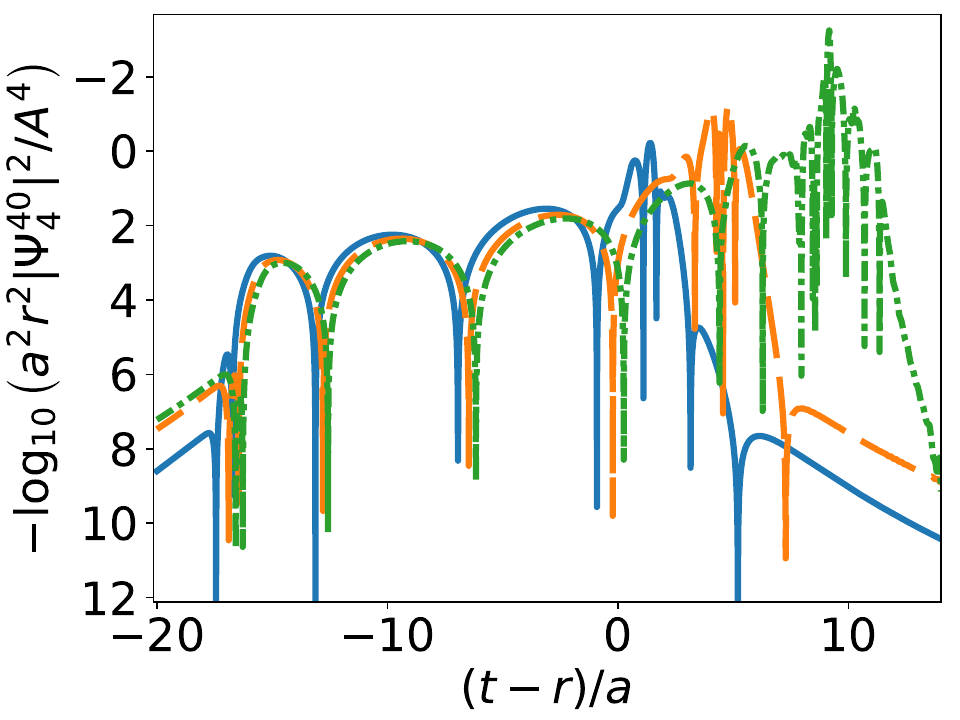}
  \includegraphics[width=0.32\linewidth]{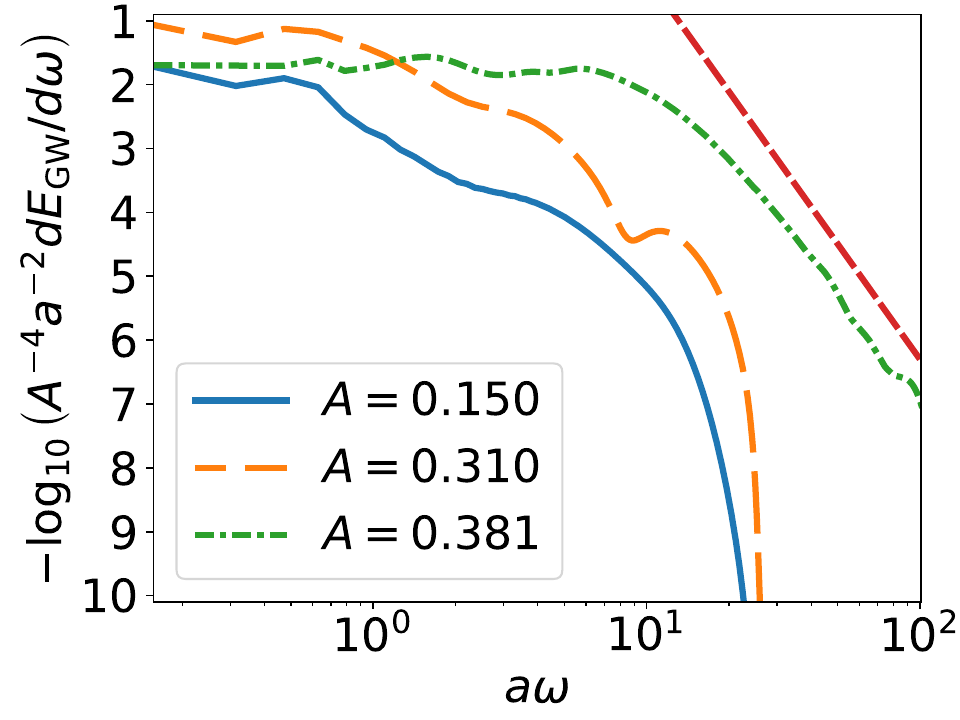}
  \caption{
  \label{fig:comparison_gws_headon}
  Radiation in head-on pulse collisions for different amplitudes,
  extracted at $r=90a$.
  {\bf Left:} EM radiation.
  {\bf Middle:} Mode $(l, m) = (4, 0)$ of $\Psi_4$.
  {\bf Right:} Power spectrum of gravitational wave emission,
  from all of the modes up to $l=12$.
  As a reference, we also show
  a best fit power law of the form $\omega^{-6}$.
  }
\end{figure*}

Fig.~\ref{fig:comparison_gws_headon} shows the EM and gravitational
radiation of subcritical solutions.
The behavior is similar to the single pulse case,
with a more pronounced time delay in the maxima as the
threshold of collapse is approached.
The ringdown frequencies are significantly higher than for the single pulse
case,
which implies that the emitting compact state has a smaller extent.
From the eikonal approximation,
we estimate the radius of the light ring of the initial compact
state to be $r_\mathrm{light} = 0.64a$ for the simulation with $A=0.381$.
This fits with the observation that,
for the amplitudes considered,
both the ADM mass contribution from each pulse and their compactness,
which dictate the subset of the mass that contributes to the ringdown,
are lower in the head-on case.
Nonlinearities also seem to create structure at small scales,
which is apparent in the frequency domain.
Figure~\ref{fig:comparison_gws_headon} shows the power spectrum of the GW
signal,
for a hexadecapolar ($l=4$) component.
For low amplitudes,
the spectrum decays sharply for $\omega a\gtrsim 1$.
However, close to threshold,
the signal contains appreciable amount of energy at high frequencies,
indicating a transfer of energy to smaller scales.
Indeed, Fig.~\ref{fig:double_ringdown} shows that,
after the peak,
all modes have similar amplitudes,
indicating that the nonlinearities result in an excitation of the smaller
scales.

\noindent\textbf{\emph{Discussion.}}
We have shown that strong pulses of light interacting gravitationally present
rich dynamics and until now unexplored phenomenology.
By evolving the Einstein-Maxwell system fully nonlinearly, we have found such
effects as ultracompact self-gravitating states, double ringdown,
and strong transfer of energy to small scales.
This study opens the possibility of studying ultrarelativistic collisions by
considering massless fields.

The physics of these collisions are extreme,
for example the measured luminosities are close to the Dyson-Thorne
bound~\cite{Dys63,Tho83,Sperhake:2012me,Car13}.
The compactness measured by Eulerian observers in the nonlinear regime
surpasses the proposed maximum given by the hoop conjecture,
resulting in very compact self-gravitating electromagnetic states.
The relation between such compact configurations, their lifetime and Wheeler's
geons~\cite{Wheeler:1955zz} will be reported elsewhere.

Recently it has been claimed~\cite{AlvGarMar24} that the Schwinger effect would
preclude BH formation from electromagnetic radiation, although doubts have been
shed on the generality of the result~\cite{Page:2025ypy}.
This question is out of scope for the present work,
which concerns classical GR,
and where quantum effects are ignored.
In any case,
the scale of our system can always be chosen such that the electric field is
below the Schwinger limit, increasing the mass of the formed BHs,
consistent with the analysis of~\cite{AlvGarMar24}.
In future work,
the relevance of the Schwinger effect for small BH formation
should be evaluated from the magnitude of the electromagnetic invariant
$F_{\mu\nu} F^{\mu\nu} = 2 (B^2 - E^2)$.

In relation with the picture of critical collapse,
our results indicate that the physics of solutions close to the threshold of
collapse are dictated by the presence of compact states,
even in the absence of apparent horizons.

\noindent\textbf{\emph{Acknowledgments.}}
We thank Nicola Franchini, David Hilditch, Jaime Redondo Yuste, Diogo Ribeiro,
João Santos, Leart Sabani and Lucía Vélez Tartajo for helpful comments and
discussion.
The Center of Gravity is a Center of Excellence funded by the Danish National Research Foundation under grant No.\@ DNRF184.
We acknowledge support by VILLUM Foundation (grant no.\@ VIL37766).
V.\,C.\@ is a Villum Investigator.  
H.\,R.\,R.\@ and V.\,C.\@ acknowledge financial support provided under the European Union’s H2020 ERC Advanced Grant ``Black holes: gravitational engines of discovery'' grant agreement no.\@ Gravitas-101052587.
Views and opinions expressed are, however, those of the authors only and do not necessarily reflect those of the European Union or the European Research Council. Neither the European Union nor the granting authority can be held responsible for them.
This project has received funding from the European Union's Horizon 2020 research and innovation programme under the Marie Sk{\l}odowska-Curie grant agreement No 101007855 and No 101131233.
This work is supported by Simons Foundation International and the Simons Foundation through Simons Foundation grant SFI-MPS-BH-00012593-11.
The authors gratefully acknowledge the Gauss Centre for Supercomputing
e.V.~(www.gauss-centre.eu) for funding this project by providing
computing time on the GCS Supercomputer SuperMUC-NG at Leibniz
Supercomputing Centre (www.lrz.de).
The Tycho supercomputer hosted at the SCIENCE HPC center at the University of
Copenhagen was used for supporting this work.
We acknowledge the Minho Advanced Computing Center (MACC) for providing
computing time on the Deucalion supercomputer, with FCT/FCCN project numbers
2025.09520.CPCA.A3 and 2025.09510.CPCA.A1.

\noindent\textbf{\emph{Data availability.}}
The simulation parameter files, gravitational waveforms, EM luminosity curves and compactness data generated in
this study are available on Zenodo~\cite{ZenodoData}.

\bibliography{draft.bbl}

\clearpage
\newpage

\appendix
\section*{END MATTER}
\section{Horizon properties}
\label{section:apparenthorizons}

\begin{table}
 \begin{tabular}{ccccccc}
 \toprule
  $A$ & $t_{\rm AH}/a$ & $z_K/a$ & $m_{\rm AH}/a$ & $\dot{m}_{\rm AH}$ & $M_{\rm ADM}/a$
      & $m_{\rm AH}/M_{\rm ADM}$\\
  \midrule
  0.528& 4.84 & 3.78  & 0.0564  & 1.0 & 0.277 & 0.20\\
  0.6 &  0.51 & 1.34  & 0.0989  & 4.9 & 0.355 & 0.54\\
  0.7 & -0.40 & 1.21  & 0.171   & 3.5 & 0.484 & 0.35\\
  0.8 & -0.92 & 0.978 & 0.227   & 7.8 & 0.630 & 0.36\\
  0.9 & -1.29 & 0.961 & 0.351   & 6.9 & 0.795 & 0.44\\
  1.0 & -1.68 & 0.940 & 0.502   & 16  & 0.979  & 0.51\\
  \bottomrule
 \end{tabular}
 \caption{Properties of the first detected
 apparent horizon in single pulse simulations. The horizon first forms at
 $t_{\rm AH}$,
 $m_{\rm AH}$ is its horizon mass,
 and $\dot{m}_{\rm AH}$ the accretion rate of the BH.
 $z_K$ is the position of the maximum of the Kretschmann scalar at $t_{\rm AH}$,
 the prospective singularity point.
 $M_{\rm ADM}$ is the ADM mass measured in the initial data.\label{tab:pulse_horizons}
 }
\end{table}

\begin{table}
 \begin{tabular}{ccccccc}
 \toprule
  $A$ & $t_{\rm AH}/a$ & $m_{\rm AH}/a$ & $M_{\rm ADM}/a$
      & $m_{\rm AH}/M_{\rm ADM}$\\
  \midrule
  0.384 & 4.61  & 0.122 & 0.287 & 0.42\\
  0.4 & 2.36    & 0.275 & 0.319 & 0.86\\
  0.5 & 0.18    & 0.300 & 0.499 & 0.60\\
  0.6 & -0.62   & 0.458 & 0.716 & 0.64\\
  0.7 & -1.24   & 0.694 & 0.972 & 0.71\\
  0.8 & -1.91   & 0.895 & 1.27  & 0.70\\
  0.9 & -2.53   & 1.09  & 1.60  & 0.68\\
  \bottomrule
 \end{tabular}
\caption{
  \label{tab:headon_horizons} Properties of the first detected apparent horizon
  in head-on collisions, for different amplitudes $A$.
  $t_{\rm AH}$ is the time at which the first horizon is found
  and $m_{\rm AH}$ is its horizon mass.
  $M_{\rm ADM}$ is the ADM mass measured in the initial data.
  }
\end{table}

Tables~\ref{tab:pulse_horizons} and~\ref{tab:headon_horizons} list the apparent horizon properties at the
moment the horizon finder detects them for the first time in single pulse
and head-on collision simulations respectively.
Table~\ref{tab:pulse_horizons} gives the accretion rates around the time of
horizon formation.
The accretion rates are of order 1, which is higher than any known accretion
rates in astrophysical scenarios.
For comparison,
the accretion rate of the SgrA source is $\sim 3\times
10^{-20}$~\cite{2007ApJ...654L..57M},
and the maximum accretion rates that can be reconstructed from the
neutron star merger simulations of~\cite{ShiTan06}
are on the order of $10^{-3}$.

\section{Comparison with eikonal approximation}
\label{section:eikonal}
\begin{figure*}
  \centering
  \includegraphics[width=0.49\linewidth]{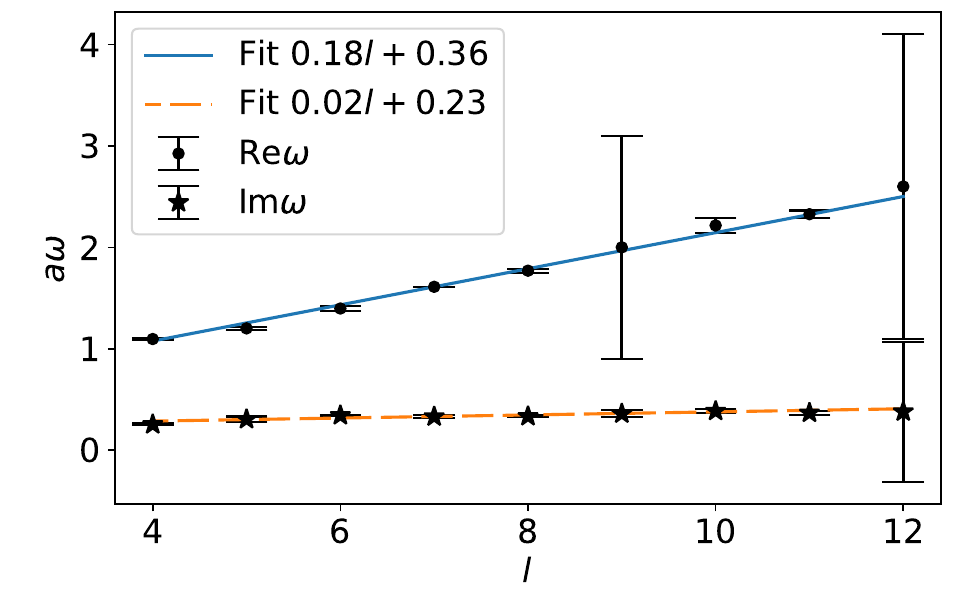}
  \includegraphics[width=0.49\linewidth]{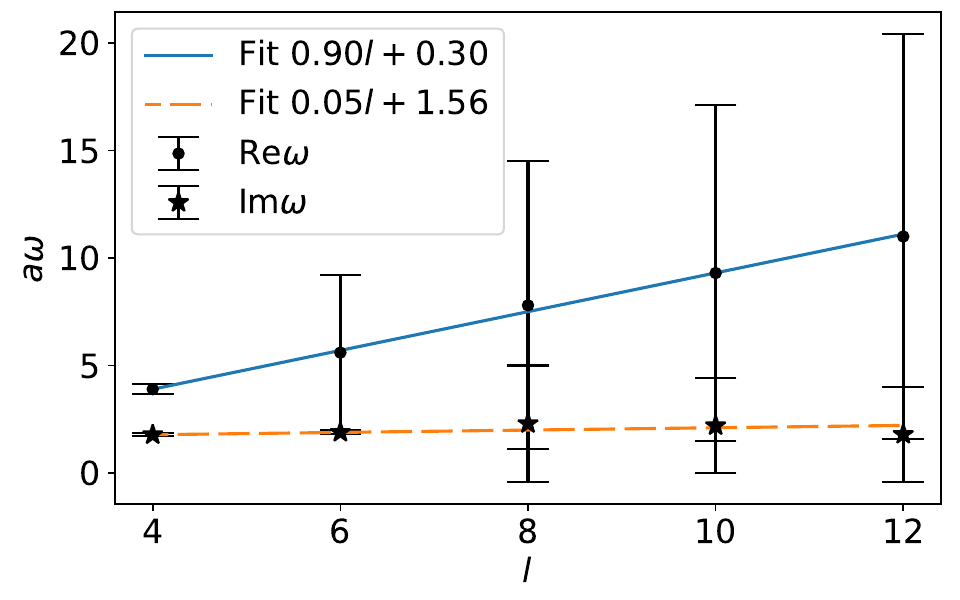}
  \caption{
  \label{fig:eikonal_fit}
  Fit of the measured ringdown frequencies to the eikonal approximation,
  see equation~\eqref{eq:eikonal}.
  {\bf Left:} Single pulse simulation with amplitude 0.5.
  {\bf Right:} First ringdown of pulse head-on collision simulation with
  amplitude 0.381.
  The measured frequencies conform well to that approximation,
  supporting our interpretation of the signals as ringdown.
  }
\end{figure*}

The eikonal approximation (see~\cite{Berti:2009kk}) gives the following behavior for high $l$ modes
\begin{gather}
  \omega \simeq \Omega_c l - i (n + 1/2) \lambda\,,
  \label{eq:eikonal}
\end{gather}
$\Omega_c$ and $\lambda$ are constants that determine the orbital frequency and
Lyapunov exponent of close-to-light-ring null geodesics.
For Schwarzschild BHs,
this approximation provides good results up to relatively low values of
$l$~\cite{Berti:2009kk}.
$n$ is the overtone number.
In Fig.~\ref{fig:eikonal_fit}, we fit the frequencies of several modes in two
representative simulations to the eikonal expectation.
We find good agreement, which supports the idea that what we are seeing is
ringdown,
even if the spacetime is not stationary.

In a Schwarzschild background,
the orbital frequency of the light-ring is related to the mass of the BH
as~\cite{Press:1971wr}
\begin{gather}
  \Omega_c = \frac{1}{3\sqrt3 M}\,,
\end{gather}
and the light-ring is located at a radius of $r_\mathrm{light} = 3M$,
so that $r_\mathrm{light} = \left(\sqrt3 \Omega_c\right)^{-1}$.
We can use this relation to estimate the radius of an appropriate generalization of the light-ring four our dynamical, horizonless compact state.

The quality factors of the modes shown in Fig.~\ref{fig:eikonal_fit} are smaller
than in a Schwarzschild BH.
For horizonless compact configurations,
quality factors are usually higher, because energy is no longer lost through the horizon (for example, quality factors of the f-mode of neutron stars can be ${\cal O} (10^3)$~\cite{Kokkotas:1999bd}).
In our case,
there is a sub-region of spacetime with very high compactness,
but there is also nonzero energy density everywhere else too.
The presence of matter around the light-ring might make the null geodesics
around the light-ring more unstable.

The section of the signal that shows ringdown can be very short in some cases,
particularly close to the threshold for the head-on collisions of EM pulses.
This introduces a dependence of the fitted parameters of the chosen fitting
window that cannot be ignored.
Our method for accounting for this dependence is the following.
Consider that we have a signal given by a function $\phi_{\rm signal}$,
that has a number of points $N$.
We try different fitting windows $W$,
each of which results in a fitted signal $\phi_{\rm fit}^W$ and a series of
parameters $\{a^W_i\}_i^P$.
We choose the window that minimizes
\begin{gather}
  \frac{\norm{\phi_{\rm fit} - \phi_{\rm signal}}_{L2}}{N^p}\,.
\end{gather}
The number of points in the denominator is chosen so as to prefer larger fitting
windows,
since generally a smaller fitting window will result in smaller L2 difference
even if the fit does not improve.
In practice, we have used values of $p=2$ and $p=1$,
depending on the case.
Once the fitting window has been chosen,
the contribution to the uncertainty associated with the dependence of the
fitting window is calculated assuming a uniform distribution between the minimum
and maximum values obtained for all of the windows tested
\begin{gather}
  \sigma_i = \frac{\max_W(a_i^W) - \min_W(a_i^W)}{\sqrt{12}}\,.
\end{gather}
The final uncertainty for the fitted parameters is calculated by compounding
this uncertainty,
and the one from the nonlinear fit.

\section{Resolution comparison}
\begin{figure}
  \centering
  \includegraphics[width=0.9\linewidth]{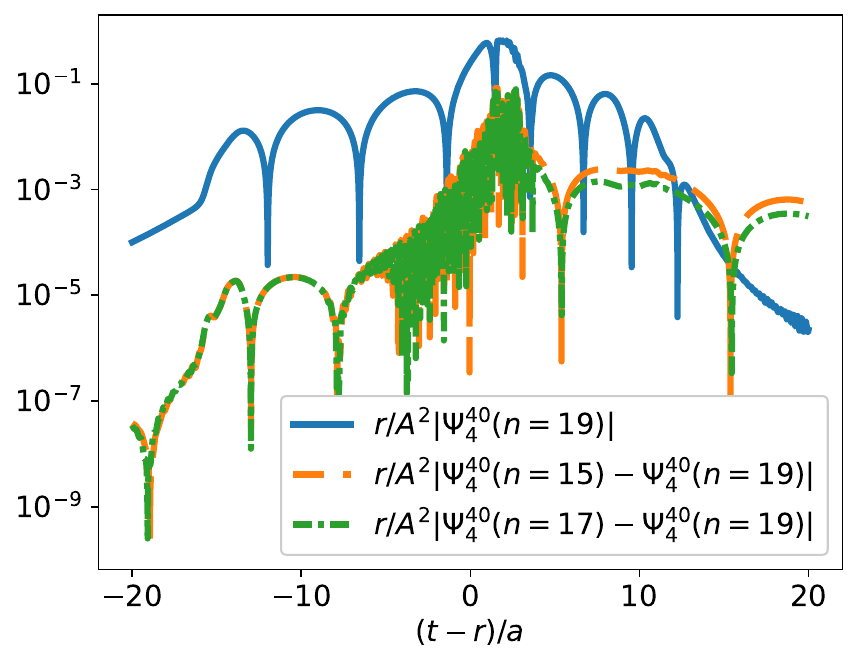}
  \caption{
  \label{fig:convergence}
  Waveform error for the $\Psi_4^{40}$ mode
  for the subcritical single pulse simulation with $A=0.5$.
  The solid line shows the absolute value of $\Psi_4^{40}$ for a simulation
  with 19 collocation points per dimension in each spectral element.
  Dashed and dash-dotted lines show the absolute value of the difference
  to the result from simulation with 15 and 17 collocation points respectively.  
  }
\end{figure}

Fig.~\ref{fig:convergence} shows a comparison of the waveforms at different 
resolutions for a typical simulation.
For the comparison,
we run a simulation with AMR and save the refinement schedule.
Then, we use the same refinement schedule for a simulation with a lower
number of collocation points per element.
From this we estimate a relative error of about 1\% at the waveform peak,
and larger at later times,
cf. Fig.~\ref{fig:convergence}.
At $t-r > 12a$ the signal is dominated by error and for this reason we exclude
it from our analysis.
Higher resolution would be required to properly resolve that section of the signal.

\section{Compactness measurements}
\label{section:compactness}
Our simulation domain is divided into numerical elements,
and we measure the compactness of coordinate ellipsoids $B(a, b)$
by finding all of the
elements $\Omega$ that intersect the ellipsoid and measuring their total mass
and volume,
\begin{gather}
  C_{B(a, b)} := \frac{2 \sum_i \int_{\Omega_i} n_\mu n_\nu T^{\mu\nu}\sqrt{\gamma}
  dV}{\left(3/4\pi \sum_i\int_{\Omega_i} \sqrt{\gamma} dV\right)^{1/3}} =
  \frac{2M}{\left(\frac{3V}{4\pi}\right)^{1/3}}\,,
\end{gather}
with $\Omega_i \cap B(a, b) \neq \emptyset$ and $\gamma = \det g_{ij}$. This amounts to measuring the compactness of a ``patchy'' coordinate ellipsoid.
Motivated by the hoop conjecture,
we choose ellipsoids such that the proper length (measured with $g_{ij}$)
in all principal directions is the same.
These are not exactly spheres in the slice geometry,
both because we are measuring over entire elements,
and we do not check the proper lengths in nonprincipal directions.

\end{document}